\pdfoutput=1
\documentclass[journal]{IEEEtran}

\usepackage[utf8]{inputenc}

\usepackage{cite}

\usepackage{booktabs}
\usepackage[pdftex]{graphicx}

\usepackage[dvipsnames]{xcolor}
\usepackage{tikz}
\usepackage{pgfplots}
\pgfplotsset{compat=1.13}

\usepackage{etoolbox}
\usepackage{microtype}

\usepackage{acro}

\usepgfplotslibrary{external}
\usepgfplotslibrary{colormaps}
\usepgfplotslibrary{units}
\usepgfplotslibrary{groupplots}
\usepgfplotslibrary{statistics}
\usetikzlibrary{chains}
\usetikzlibrary{scopes}
\usetikzlibrary{patterns}
\pgfplotsset{%
    colormap={parula}{%
        rgb255=(53,42,135)
        rgb255=(15,92,221)
        rgb255=(18,125,216)
        rgb255=(7,156,207)
        rgb255=(21,177,180)
        rgb255=(89,189,140)
        rgb255=(165,190,107)
        rgb255=(225,185,82)
        rgb255=(252,206,46)
        rgb255=(249,251,14)
    },
    colormap={hotinv}{%
        [1cm]rgb255(0cm)=(255,255,255)
        rgb255(2cm)=(255,255,0)
        rgb255(5cm)=(255,0,0)
        rgb255(8cm)=(0,0,0)
    },
}

\pgfplotsset{%
    every x tick label/.append style={/pgf/number format/1000 sep=},
    unit markings=slash space,
    evalline/.style={line width=1pt, mark=*, mark size=1pt, mark options={solid}},
    log_noisy/.style={dashed, blue, evalline},
    log_spp_tcs_prio/.style={densely dotted, red, evalline},
    generic/.style={black, evalline},
    omlsa/.style={gray, evalline},
    log_pery/.style={dashed, OliveGreen, evalline},
    log_pery_psdn/.style={dashed, SpringGreen, evalline},
    log_apost/.style={densely dotted, SkyBlue, evalline},
    log_aprior/.style={densely dotted, Blue, evalline},
    log_apost_aprior/.style={dashdotdotted, Magenta, evalline},
    noisy/.style={dashdotdotted, Orange, evalline},
    onelayer/.style={opacity=0.4},
}
\input{colorconverter}

\newenvironment{customlegend}[1][]{%
    \begingroup
    \csname pgfplots@init@cleared@structures\endcsname
    \pgfplotsset{#1}%
}{%
    \csname pgfplots@createlegend\endcsname
    \endgroup
}%

\def\addlegendimage{\csname pgfplots@addlegendimage\endcsname}

\usepackage{amsmath}
\usepackage{amssymb}

\usepackage{algorithm}
\usepackage{algorithmic}

\usepackage{array}

\ifCLASSOPTIONcompsoc
  \usepackage[caption=false,font=normalsize,labelfont=sf,textfont=sf]{subfig}
\else
  \usepackage[caption=false,font=footnotesize]{subfig}
\fi

\usepackage{url}

\DeclareAcronym{ANOVA}{short=ANOVA, long=analysis of variance}
\DeclareAcronym{BECOCO}{short=BECOCO, long=phase-(b)lind (e)stimator of (co)mplex (co)efficients}
\DeclareAcronym{CDF}{short=CDF, long=cumulative distribution function}
\DeclareAcronym{CMVN}{short=CMVN, long=cepstral mean and variance normalization}
\DeclareAcronym{CSNE}{short=CSNE, long=concatenated short noise excerpts, long-plural=concatenated short noise excerpts}
\DeclareAcronym{DFT}{short=DFT, long=discrete Fourier transform}
\DeclareAcronym{DNN}{short=DNN, long=deep neural network}
\DeclareAcronym{GMM}{short=GMM, long=Gaussian mixture model}
\DeclareAcronym{HMM}{short=HMM, long=hidden Markov model}
\DeclareAcronym{IDFT}{short=IDFT, long=inverse discrete Fourier transform}
\DeclareAcronym{IRM}{short=IRM, long=ideal ratio mask}
\DeclareAcronym{IS}{short=IS, long=Itakura-Saito}
\DeclareAcronym{MFCC}{short=MFCC, long=Mel-frequency cepstral coefficient}
\DeclareAcronym{MLSE}{short=MLSE, long=machine-learning spectral envelope}
\DeclareAcronym{ML}{short=ML, long=machine-learning}
\DeclareAcronym{MMSE}{short=MMSE, long=minimum mean-squared error}
\DeclareAcronym{MOSIE}{short=MOSIE, long=(M)MSE estimation with (o)ptimizable (s)peech (m)odel and (i)nhomogeneous (e)rror criterion}
\DeclareAcronym{MSE}{short=MSE, long=mean-squared error}
\DeclareAcronym{MUSHRA}{short=MUSHRA, long=multi-stimulus test with hidden reference and anchor}
\DeclareAcronym{MoG}{short=MoG, long=mixture of Gaussians, long-plural-form=mixtures of Gaussians}
\DeclareAcronym{NMF}{short=NMF, long=nonnegative matrix factorization}
\DeclareAcronym{OMLSA}{short=OMLSA, long=optimally modified log-spectral amplitude estimator}
\DeclareAcronym{PDF}{short=PDF, long=probability density function}
\DeclareAcronym{PESQ}{short=PESQ, long=Perceptual Evaluation of Speech Quality}
\DeclareAcronym{PSD}{short=PSD, long=power spectral density}
\DeclareAcronym{ReLU}{short=ReLU, long=rectified linear unit}
\DeclareAcronym{SNR}{short=SNR, long=signal-to-noise ratio}
\DeclareAcronym{SPP}{short=SPP, long=speech presence probability}
\DeclareAcronym{STFT}{short=STFT, long=short-time Fourier transform} \DeclareAcronym{VTS}{short=VTS, long=vector Taylor series} \DeclareAcronym{LSA}{short=LSA, long=log-spectral amplitude estimator}
\DeclareAcronym{STOI}{short=STOI, long=short-time objective intelligibility}
\DeclareAcronym{STSA}{short=STSA, long=short-term spectral amplitude estimator}
\DeclareAcronym{SegNR}{short=SegNR, long=segmental noise reduction}
\DeclareAcronym{SegSSNR}{short=SegSSNR, long=segmental speech \ac{SNR}}
\DeclareAcronym{TCS}{short=TCS, long=temporal cepstrum smoothing}

\newcommand{\sFreqIdx}{k}
\newcommand{\sFrameIdx}{\ell}

\newcommand{\sCepsIdx}{q}
\newcommand{\sIdxPair}{{\sFreqIdx, \sFrameIdx}}
\newcommand{\sCIdxPair}{{\sCepsIdx, \sFrameIdx}}

\newcommand{\sNoise}{n}
\newcommand{\sSpeech}{s}
\newcommand{\sNoisy}{y}

\newcommand{\sFreq}[1]{\MakeUppercase{#1}}

\newcommand{\sLog}[1]{{#1}^\text{(log)}}

\newcommand{\sNoiseF}{\sFreq{\sNoise}}
\newcommand{\sNoiseFI}{\sNoiseF_{\sIdxPair}}
\newcommand{\sNoiseFIEst}{\sEst{\sNoiseF}_{\sIdxPair}}

\newcommand{\sSpeechF}{\sFreq{\sSpeech}}
\newcommand{\sSpeechFIEst}{\sEst{\sSpeechF}_{\sIdxPair}}
\newcommand{\sSpeechFI}{\sSpeechF_{\sIdxPair}}

\newcommand{\sNoisyF}{\sFreq{\sNoisy}}
\newcommand{\sNoisyFI}{\sNoisyF_{\sIdxPair}}

\newcommand{\sNoisyLog}{\sLog{\sNoisy}}
\newcommand{\sNoisyLogI}{\sNoisyLog_{\sIdxPair}}

\newcommand{\sEst}[1]{\hat{#1}}

\newcommand{\sCovF}{\Lambda}

\newcommand{\sVarNoiseF}{\sCovF^{\sNoise}_\sIdxPair}
\newcommand{\sVarNoiseFLog}{\sCovF^{\sNoise, \text{(log)}}_\sIdxPair}
\newcommand{\sVarSpeechF}{\sCovF^{\sSpeech}_\sIdxPair}

\newcommand{\sVarNoiseFEst}{\sEst{\sCovF}^{\sNoise}_\sIdxPair}
\newcommand{\sVarNoiseFEstLog}{\sEst{\sCovF}^{\sNoise, \text{(log)}}_\sIdxPair}
\newcommand{\sVarNoiseFEstP}{\sEst{\sCovF}^{\sNoise}_{\sFreqIdx, \sFrameIdx-1}}
\newcommand{\sVarSpeechFEst}{\sEst{\sCovF}^{\sSpeech}_\sIdxPair}
\newcommand{\sVarSpeechFEstML}{\sEst{\sCovF}^{\sSpeech, \text{ml}}_\sIdxPair}
\newcommand{\sVarSpeechCEstML}{\sEst{\sCovF}^{\sSpeech, \text{ml}}_\sCIdxPair}
\newcommand{\sVarSpeechCEst}{\sEst{\sCovF}^{\sSpeech}_\sCIdxPair}
\newcommand{\sVarSpeechCEstP}{\sEst{\sCovF}^{\sSpeech}_{\sCepsIdx, \sFrameIdx-1}}

\newcommand{\sProb}{P}
\newcommand{\sPDF}{p}

\newcommand{\sGain}{G}
\newcommand{\sGainI}{G_{\sIdxPair}}

\newcommand{\sGainMin}{\sGain_{\text{min}}}

\newcommand{\sPriorSNR}{\xi}
\newcommand{\sPriorSNRI}{\sPriorSNR_{\sIdxPair}}
\newcommand{\sPriorSNRILog}{\sPriorSNR^\text{(log)}_{\sIdxPair}}

\newcommand{\sPriorSNRMinML}{\sPriorSNR^\text{ml}_\text{min}}
\newcommand{\sPostSNR}{\gamma}
\newcommand{\sPostSNRI}{\sPostSNR_{\sIdxPair}}
\newcommand{\sPostSNRILog}{\sPostSNR^\text{(log)}_{\sIdxPair}}

\newcommand{\sBias}{\varepsilon}

\newcommand{\sCost}{J}

\newcommand{\sIRM}{\text{IRM}}
\newcommand{\sIRMEst}{\widehat{\sIRM}}

\newcommand{\sHypothesis}{H}
\newcommand{\sHypoSpeech}{\sHypothesis_1}
\newcommand{\sHypoNoise}{\sHypothesis_0}
\newcommand{\sSNROpt}{\xi_{\sHypoSpeech}}
\newcommand{\sSmoothFix}{\beta}

\newcommand{\sSmoothCI}{\alpha_\sCIdxPair}
\newcommand{\sBiasC}{\kappa}

\begin{document}
\bstctlcite{IEEEexample:BSTcontrol}

\title{Normalized Features for Improving the Generalization of DNN Based Speech Enhancement}

\author{Robert~Rehr~and
        Timo~Gerkmann
}

\markboth{}%
{}

\maketitle

\begin{abstract}
    Enhancing noisy speech is an important task to restore its quality and to improve its intelligibility.
    In traditional non-\ac{ML} based approaches the parameters required for noise reduction are estimated blindly from the noisy observation while the actual filter functions are derived analytically based on statistical assumptions.
    Even though such approaches generalize well to many different acoustic conditions, the noise suppression capability in transient noises is low.
    To amend this shortcoming, machine-learning (\ac{ML}) methods such as deep learning have been employed for speech enhancement.
    However, due to their data-driven nature, the generalization of \ac{ML} based approaches to unknown noise types is still discussed.
    To improve the generalization of \ac{ML} based algorithms and to enhance the noise suppression of non-\ac{ML} based methods, we propose a combination of both approaches.
    For this, we employ the \emph{a priori} \ac{SNR} and the \emph{a posteriori} \ac{SNR} estimated as input features in a \ac{DNN} based enhancement scheme.
    We show that this approach allows \ac{ML} based speech estimators to generalize quickly to unknown noise types even if only few noise conditions have been seen during training.
    Further, the proposed features outperform a competing approach where an estimate of the noise power spectral density is appended to the noisy spectra.
    Instrumental measures such as \ac{PESQ} and \ac{STOI} indicate strong improvements in unseen conditions when the proposed features are used.
    Listening experiments confirm the improved generalization of our proposed combination.
\end{abstract}


\IEEEpeerreviewmaketitle%

\acresetall%

\section{Introduction}
\IEEEPARstart{I}{n} the presence of background noise, speech may be distorted such that the speech intelligibility, as well as the quality of the speech signal is deteriorated.
Besides human perception, background noise also affects automatic speech recognition algorithms for human-machine interfaces and results in lower recognition rates.
Speech enhancement algorithms therefore play an important role for noise robust speech recognition and for improving the speech quality in hearing aid and telecommunication applications.
In this paper, single-channel speech enhancement algorithms are considered that either assume that the noisy signal has been captured by a single microphone or process the output of a beamformer~\cite{vary_digital_2006}.

Single-channel speech enhancement has been a research topic for many decades and has led to many different methods, e.g.,~\cite{ephraim_speech_1984, ephraim_bayesian_1992, cohen_speech_2001, cohen_noise_2003, srinivasan_codebook_2006, zhao_hmm-based_2007, breithaupt_novel_2008, gerkmann_noise_2011, gerkmann_unbiased_2012, mohammadiha_supervised_2013, xu_regression_2015, chazan_hybrid_2016}.
Here, we distinguish between two broad categories of single-channel speech enhancement schemes, namely \ac{ML} based and non-\ac{ML} based approaches.
\ac{ML} based enhancement schemes generally follow a two-step approach to enhance a noisy speech signal.
First, the model parameters of an \ac{ML} algorithm are tuned on training examples.
After that, the obtained models are used to separate the speech component from the background noise.
On the contrary, non-\ac{ML} based approaches do not learn any models from training data prior to processing.
Instead carefully designed algorithms are used to estimate the parameters required for the enhancement on-line and blindly from the noisy observation.

Non-\ac{ML} based enhancement schemes such as~\cite{ephraim_speech_1984, ephraim_speech_1985, martin_noise_2001, breithaupt_novel_2008, gerkmann_noise_2011,gerkmann_unbiased_2012} commonly operate in the \ac{STFT} domain where a filter function is applied to suppress the coefficients that mainly contain noise.
The employed filter functions are often derived in a statistical framework where the speech and noise coefficients are modeled by parametric distributions.
The consideration of various statistical models and target functions has led to many different solutions~\cite{ephraim_speech_1984, ephraim_speech_1985, breithaupt_parameterized_2008, hendriks_dft-domain_2013}.
The estimators are functions of the distributions' parameters such as the speech \ac{PSD} and the noise \ac{PSD}, which are estimated blindly from the noisy observation.
Various methods have been proposed to estimate the noise \ac{PSD}, e.g.~\cite{martin_noise_2001, cohen_speech_2001, cohen_noise_2003, gerkmann_noise_2011, gerkmann_unbiased_2012}.
Generally, the algorithms' design is based on the assumption that the background noise changes more slowly than the speech signal.
From the noise \ac{PSD} estimate and the noisy observation, the speech \ac{PSD} is obtained, e.g.,~\cite{ephraim_speech_1984, breithaupt_novel_2008}.
Non-\ac{ML} approaches have been proven to generalize well to many different acoustical environments and provide good results in moderately varying noise types.
However, they lack the ability to track fast changes of the background noise due to their underlying assumptions.
As a consequence, transient sounds such as the cutlery in a restaurant environment are generally not suppressed by non-\ac{ML} based enhancement schemes.

The shortcomings of non-\ac{ML} based enhancement schemes especially with respect to the limited noise tracking capabilities have motivated the usage of \ac{ML} algorithms for speech enhancement, e.g.,~\cite{ephraim_bayesian_1992, burshtein_speech_2002, srinivasan_codebook_2006, zhao_hmm-based_2007, mohammadiha_supervised_2013, wang_training_2014, xu_regression_2015, chazan_hybrid_2016, chen_large-scale_2016, he_multiplicative_2017, kolbaek_speech_2017}.
For this, various \ac{ML} algorithms have been considered, e.g., codebooks~\cite{srinivasan_codebook_2006, he_multiplicative_2017}, hidden Markov models~\cite{ephraim_bayesian_1992, zhao_hmm-based_2007}, Gaussian mixture models~\cite{burshtein_speech_2002} and non-negative matrix factorization~\cite{mohammadiha_supervised_2013}.
Recently, \acp{DNN} are more intensively investigated for speech enhancement applications~\cite{wang_training_2014, xu_regression_2015, chazan_hybrid_2016, chen_large-scale_2016, kolbaek_speech_2017}.
Neural networks potentially allow to approximate any non-linear function on a limited range of the input space.
They have been used to replace or to improve building blocks, e.g., the speech \ac{PSD} and noise \ac{PSD} estimation in non-\ac{ML} algorithms~\cite{suhadi_data-driven_2011, xia_wiener_2014, chinaev_noise-presence-probability-based_2016}.
Other approaches use \acp{DNN} to find a mapping from the noisy observation or features extracted from it to a filter or masking function~\cite{wang_training_2014, chen_large-scale_2016, kolbaek_speech_2017}.
\Acp{DNN} can as well be utilized to learn a mapping where the target is directly given by the clean speech coefficients as in~\cite{weninger_discriminatively_2014, xu_regression_2015}.
For this, various network types have been considered, e.g., feed-forward networks~\cite{lu_speech_2013, xu_regression_2015}, recurrent neural networks~\cite{maas_recurrent_2012} including long short-term memory based methods~\cite{weninger_single-channel_2014}, convolutional neural networks~\cite{park_fully_2017}, generative adversarial networks~\cite{pascual_segan:_2017, michelsanti_conditional_2017} and WaveNet based architectures~\cite{van_den_oord_wavenet:_2016, qian_speech_2017}.
Studies on \ac{ML} enhancement approaches, e.g.,~\cite{xu_regression_2015, chen_large-scale_2016, kolbaek_speech_2017}, show that \ac{DNN} based approaches in principal have the ability to reduce transient noises, but one of the major concerns towards \ac{ML} based approaches is their generalization to noise types that have not been seen during training.
This issue is encountered for example with large and diverse training data~\cite{xu_regression_2015, chen_large-scale_2016}, where hundreds or even thousands of different noise types are included to allow the \ac{DNN} based enhancement schemes to generalize to unseen noise conditions.
Even though large training sets increase the generalization, a huge number of noise types may still be inappropriate as in real scenarios virtually infinitely many noise types can possibly occur as argued in~\cite{chazan_hybrid_2016, kumar_speech_2016}.

To improve the robustness in unseen noise conditions, other approaches incorporate estimates of non-\ac{ML} based noise \ac{PSD} estimators, e.g.,~\cite{seltzer_investigation_2013, xu_dynamic_2014, xu_regression_2015, kumar_speech_2016}.
For this, an estimate of the noise \ac{PSD} is appended to the noisy input features which is referred to as noise aware training.
In~\cite{seltzer_investigation_2013, xu_regression_2015}, a fixed noise \ac{PSD} estimate is used which has been obtained from the first frames of the noisy input signal.
In~\cite{xu_dynamic_2014, kumar_speech_2016}, this idea has been advanced by employing a dynamic, i.e., time-varying noise \ac{PSD} estimate, obtained from a non-\ac{ML} based estimator.
However, the results in~\cite{xu_dynamic_2014, kumar_speech_2016} show only small improvements over the approaches that are not aware of the background noise if a non-\ac{ML} approach is used to estimate the noise \ac{PSD}.

In this paper, we show that non-\ac{ML} based estimates of the speech and noise \ac{PSD} can considerably increase the robustness of \ac{ML} based speech enhancement schemes towards unseen noises.
In contrast to noise aware training approaches, we propose to employ the \emph{a priori} \ac{SNR}, i.e., the ratio between the noisy periodogram and the noise \ac{PSD}, and the \emph{a posteriori} \ac{SNR}, i.e., the ratio of the noisy periodogram and the noise \ac{PSD}, as features.
Thus, instead of appending the noise \ac{PSD} to the input features extracted from the noise observation as in~\cite{seltzer_investigation_2013, xu_dynamic_2014, xu_regression_2015, kumar_speech_2016}, here, the noise \ac{PSD} estimate is used for normalization.
The usage of the \emph{a priori} \ac{SNR} and the \emph{a posteriori} \ac{SNR} is motivated by non-linear clean speech estimators, e.g.,~\cite{ephraim_speech_1984, ephraim_speech_1985} where these quantities result from the derivation of Bayesian estimators.
We show that the proposed features outperform features where the noise \ac{PSD} estimate is appended to the noisy input vector.
Further, the proposed \ac{SNR} based features have the advantage that the enhancement system is independent of the scaling of the input signal, i.e., the overall level has no effect on the enhancement.

These claims are confirmed in the evaluation using instrumental measures.
For this, \ac{PESQ}~\cite{noauthor_p.862:_2001} scores and the \ac{STOI}~\cite{taal_algorithm_2011} are evaluated in a cross-validation based experimental setup, where different sets of noise types for training and testing are used.
Further, the instrumental evaluation is supported by subjective evaluations.
First, we describe the employed algorithms in Section~\ref{sec:NonMLAlgorithms} and Section~\ref{sec:DNN}.
The results of the instrumental evaluation is given in Section~\ref{sec:InstrumentalEvaluation} while the subjective evaluation is described in Section~\ref{sec:SubjectiveEvaluation}.

\section{Non-\acs{ML} Enhancement Algorithms}\label{sec:NonMLAlgorithms}

This section gives an overview over the non-\ac{ML} based enhancement algorithms which form the basis of the proposed features in Section~\ref{sec:DNN}.
First, the non-\ac{ML} based estimation of the clean speech coefficients is considered which is based on the Wiener filter.
After that, the non-\ac{ML} based speech and noise \ac{PSD} estimators used in this paper are described.

\subsection{Estimation of the Clean Speech Coefficients}

For estimating the clean speech coefficients, the non-\ac{ML} based clean speech estimators makes use of the \ac{STFT}.
This representation is obtained by splitting the noisy input signal into overlapping segments and taking the Fourier transform of each segment after a tapered spectral analysis window has been applied.
The physically plausible assumption is used that the speech signal and the noise signal mix additively, i.e.,
\begin{equation}
    \label{eq:SignalModel}
    \sNoisyFI = \sSpeechFI + \sNoiseFI.
\end{equation}
The symbols $\sSpeechFI$ and $\sNoiseFI$ denote the complex clean speech spectrum and the complex noise spectrum, respectively, while $\sNoisyFI$ is the resulting spectrum of the noisy signal.
Furthermore, $\sFreqIdx$ is the frequency index and $\sFrameIdx$ is the segment index.
The speech coefficients are estimated using the Wiener filter gain function $\sGainI$ as
\begin{equation}
    \label{eq:GainApplication}
    \sSpeechFIEst = \max(\sGainI, \sGainMin) \sNoisyFI,
\end{equation}
where $\sGainMin$ acts as a lower limit on the Wiener filter gain~$\sGainI$.
The minimum gain~$\sGainMin$ is an important parameter to limit artifacts in the enhanced signal such as fluctuations in the reduced background noise or musical tones~\cite{berouti_enhancement_1979}.
The Wiener gain $\sGainI$ is given by
\begin{equation}
    \label{eq:WienerFilter}
    \sGainI = \frac{\sVarSpeechF}{\sVarSpeechF + \sVarNoiseF}, 
\end{equation}
where $\sVarSpeechF$ and $\sVarNoiseF$ denote the speech \ac{PSD} and the noise \ac{PSD}, respectively.
The clean speech estimates $\sSpeechFIEst$ are transformed back to the time-domain for each segment~$\sFrameIdx$.
Each enhanced time-domain segment is weighted by a tapered synthesis window and by using an overlap-add method, the time-domain signal is reconstructed.

The Wiener filter is the \ac{MMSE} optimal estimator of the clean speech coefficients if the speech coefficients~$\sSpeechFI$ and the noise coefficients~$\sNoiseFI$ are assumed to be uncorrelated and to follow a complex circular-symmetric Gaussian distribution.
This assumption is often justified by the central limit theorem which may be argued to apply due to the Fourier sum which needs to be evaluated for obtaining the spectral coefficients~\cite[Chapter~4]{brillinger_time_2001}.
The speech \ac{PSD}~$\sVarSpeechF$ and the noise \ac{PSD}~$\sVarNoiseF$ are estimated blindly from the noisy observation using~\cite{gerkmann_noise_2011, breithaupt_novel_2008}.
Both algorithms are summarized in the following sections.

\subsection{Non-\acs{ML} Noise \acs{PSD} Estimation}%
\label{sec:NonMLAlgorithms:NoisePSD}

The algorithm presented in~\cite{gerkmann_noise_2011, gerkmann_unbiased_2012} is used to estimate the noise \ac{PSD}.
This estimator allows to track moderate changes in the background noise such as passing cars.
However, it cannot track transient disturbances.
In the remainder of this section, the algorithm is briefly introduced.

The noise \ac{PSD} estimator in~\cite{gerkmann_noise_2011, gerkmann_unbiased_2012} models the complex noisy coefficients under the hypotheses of speech presence~$\sHypoSpeech$ and speech absence~$\sHypoNoise$ using parametric distributions.
Given $\sHypoNoise$, the noisy observations equals $\sNoisyFI = \sNoiseFI$ while under $\sHypoSpeech$ the noisy coefficients are given by $\sNoisyFI = \sSpeechFI + \sNoiseFI$ as in~\eqref{eq:SignalModel}.
As for the Wiener filter, the speech coefficients~$\sSpeechFI$ and the noise coefficients~$\sNoiseFI$ are assumed to follow a complex circular-symmetric Gaussian distribution.
Accordingly, the likelihoods under the hypotheses $\sHypoNoise$ and $\sHypoSpeech$, i.e.,~$\sPDF(\sNoisyFI|\sHypoNoise)$ and $\sPDF(\sNoisyFI|\sHypoSpeech)$, are also modeled using Gaussian distributions.
The \ac{SPP} is defined as the posterior probability $\sProb(\sHypoSpeech|\sNoisyFI)$ which can be obtained using Bayes' theorem.
A modified posterior is used in~\cite{gerkmann_noise_2011, gerkmann_unbiased_2012} as
\begin{equation}
    \label{eq:PosteriorDistribution}
    \sProb(\sHypoSpeech | \sNoisyFI) = {\left(1 + (1 + \sSNROpt) \exp\left(-\frac{|\sNoisyFI|^2}{\sVarNoiseFEstP} \frac{\sSNROpt}{1 + \sSNROpt}\right) \right)}^{-1},
\end{equation}
which has been derived under the assumption that the prior $\sProb(\sHypoSpeech) = \sProb(\sHypoNoise) = 1/2$.
Here, a fixed \ac{SNR}~$\sSNROpt$ is used which is interpreted as the local \ac{SNR} that is expected if the hypothesis $\sHypoSpeech$ holds~\cite{gerkmann_noise_2011, gerkmann_unbiased_2012}.
The likelihood models~$\sPDF(\sNoisyFI | \sHypoNoise)$ and~$\sPDF(\sNoisyFI | \sHypoSpeech)$ have been used to formulate a speech detection problem in~\cite{gerkmann_unbiased_2012}.
By minimizing the total risk of error~\cite{gerkmann_unbiased_2012}, the optimal value $\sSNROpt = -15~\text{dB}$ has been found.

The posterior probability $\sProb(\sHypoSpeech|\sNoisyFI)$ is used to estimate the noise periodogram as
\begin{equation}
    \label{eq:MMSENoisePeriodogram}
    |\sNoiseFIEst|^2 = (1 - \sProb(\sHypoSpeech|\sNoisyFI)) |\sNoisyFI|^2 + \sProb(\sHypoSpeech|\sNoisyFI) \sVarNoiseFEstP.
\end{equation}
The estimated noise periodogram $|\sNoiseFIEst|^2$ is smoothed temporally to obtain an estimate of the noise \ac{PSD} as
\begin{equation}
    \label{eq:SPPSmooth}
    \sVarNoiseFEst = (1 - \sSmoothFix) |\sNoiseFIEst|^2 + \sSmoothFix \sVarNoiseFEstP,
\end{equation}
where $\sSmoothFix$ is a fixed smoothing constant.
This estimator can be implemented in speech enhancement framework by evaluating~\eqref{eq:PosteriorDistribution},~\eqref{eq:MMSENoisePeriodogram} and~\eqref{eq:SPPSmooth} for each frequency band~$\sFreqIdx$ when a new segment $\sFrameIdx$ is processed.
If the noise \ac{PSD} is strongly underestimated, the \ac{SPP} in~\eqref{eq:PosteriorDistribution} is overestimated, i.e., it is close to 1.
As a result, the noise periodogram in~\eqref{eq:MMSENoisePeriodogram} may no longer be updated.
To avoid such stagnations, the \ac{SPP} is set to a lower value if it has been stuck at 1 for a longer period of time~\cite{gerkmann_noise_2011, gerkmann_unbiased_2012}.

\subsection{Non-\acs{ML} Speech \acs{PSD} Estimation}%
\label{sec:NonMLAlgorithms:SpeechPSD}

For estimating the speech \ac{PSD}~$\sVarSpeechF$, the \ac{TCS} approach described in~\cite{breithaupt_novel_2008} is employed.
In contrast to the commonly used decision-directed approach~\cite{ephraim_speech_1984}, this approach causes less isolated estimation estimation errors, which may be perceived as annoying musical tones.
In this section, we recapitulate the main concepts of this algorithm.

Under the assumption that the spectral speech and noise coefficients follow a complex circular-symmetric Gaussian distribution, the limited maximum likelihood estimator is given by~\cite{ephraim_speech_1984}
\begin{equation}
    \label{eq:MaximumLikelihood}
    \sVarSpeechFEstML = \sVarNoiseFEst \max\left(\frac{|\sNoisyFI|^2}{\sVarNoiseFEst} - 1, \sPriorSNRMinML\right),
\end{equation}
where the $\max(\cdot)$ operator in combination with~$\sPriorSNRMinML$ is used to avoid negative speech \acp{PSD} and numerical issues in the following steps.
For the practical applicability, $\sVarNoiseF$ has been replaced by its estimate $\sVarNoiseFEst$.

The maximum likelihood estimate is transformed to the cepstral domain via
\begin{equation}
    \sVarSpeechCEstML = \text{IDFT}\{\log(\sVarSpeechFEstML)\},
\end{equation}
where $\sCepsIdx$ is the quefrency index and $\text{IDFT}(\cdot)$ denotes the inverse discrete Fourier transform.
In the cepstral domain, speech can be represented by using only a few coefficients: The speech spectral envelope, which reflects the impact of the vocal tract filter, is represented by the lower coefficients with $\sCepsIdx < 2.5~\text{ms}$ whereas the speech spectral fine structure, i.e., the fundamental frequency and its harmonics, is approximated by a single peak among the high cepstral coefficients.
This peak is also referred to as pitch peak.
The compact representation of speech is exploited by the \ac{TCS} approach by using a quefrency and time dependent smoothing constant $\sSmoothCI$ to smooth $\sVarSpeechCEstML$ as
\begin{equation}
    \sVarSpeechCEst = (1 - \sSmoothCI) \sVarSpeechCEstML + \sSmoothCI \sVarSpeechCEstP.
\end{equation}
For the cepstral coefficients that are associated with speech only little smoothing is applied while the remaining cepstral coefficients are strongly smoothed.
Accordingly, $\sSmoothCI$ is set close to 0 for the lower cepstral coefficients and close to 1 for the high coefficients.
In voiced segments, the $\sSmoothCI$ in close vicinity to the cepstral pitch peak are changed to values close to 0.

The cepstrally smoothed speech \ac{PSD}~$\sVarSpeechCEst$ is transformed back to the spectral domain as
\begin{equation}
    \sVarSpeechFEst = \exp(\text{DFT}\{\sVarSpeechCEst\} + \sBiasC).
\end{equation}
As the smoothing in the cepstral domain results in a biased estimate~\cite{gerkmann_statistics_2009}, the correction term $\sBiasC$ is added.
In~\cite{breithaupt_novel_2008}, it has been argued that the bias of computing the expected value of a spectral quantity following a Gaussian distribution in the logarithmic domain amounts to the Euler constant.
Due to the smoothing, the estimate in the cepstral domain is between an instantaneous value and the expected value.
Hence, $\sBiasC$ is set $1 / 2$ of the Euler constant, i.e., $\sBiasC \approx 0.5 \cdot 0.5772\dots$ is used.
A more rigorous analysis of the bias is given in~\cite{gerkmann_statistics_2009}.

\section{\label{sec:DNN}\acs{ML} Based Enhancement Algorithms}

In this section, the \ac{ML} based algorithms used in this paper are presented.
First, the enhancement framework is described and, after that, the employed input features are considered.
Note that the algorithms share the same \ac{ML} based enhancement framework but differ in the input features.

\subsection{\label{sec:DNN:Framework}\acs{ML} Based Enhancement Framework}

The architecture of the used framework resembles the approaches that have been proposed in~\cite{xu_regression_2015, kolbaek_speech_2017}.
Similar to the non-\ac{ML} based enhancement scheme, also the \ac{ML} based approaches operate in the \ac{STFT} domain.
A feed-forward \ac{DNN} is used to predict an \ac{IRM} from the input features extracted from the noisy input signal.
The features considered in this paper are described in Section~\ref{sec:DNN:NonNormalized} and Section~\ref{sec:DNN:Normalized} in detail.
The \ac{IRM} has been proposed in~\cite{wang_training_2014} and is similar to the Wiener filter gain function shown in~\eqref{eq:WienerFilter} with the difference that the speech periodogram $|\sSpeechFI|^2$ and the noise periodogram $|\sNoiseFI|^2$ are employed instead of the respective \acp{PSD} as
\begin{equation}
    \label{eq:IRM}
    \sIRM(\sFreqIdx, \sFrameIdx) = \frac{|\sSpeechFI|^2}{|\sSpeechFI|^2 + |\sNoiseFI|^2}.
\end{equation}
Similar to the Wiener filter, the predicted \ac{IRM} obtained from the \ac{DNN} is used to estimate the clean speech coefficients~$\sSpeechFIEst$ as
\begin{equation}
    \label{eq:DNNGain}
    \sSpeechFIEst = \max\left(\sIRMEst(\sFreqIdx, \sFrameIdx), \sGainMin \right) \sNoisyFI,
\end{equation}
where $\sIRMEst(\cdot)$ denotes the \ac{IRM} estimated by the \ac{DNN}.
We enforce a lower limit $\sGainMin$ as in~\eqref{eq:GainApplication} and the time-domain signal is reconstructed using an overlap-add method.

\subsection{\label{sec:DNN:NonNormalized}Non-Normalized Features}

In this part, the non-normalized feature inputs of the \ac{DNN} are presented.
The first representative of the non-normalized features is the logarithmized noisy periodogram, i.e.,
\begin{equation}
    \label{eq:LogPer}
    \sNoisyLogI = \log\left(|\sNoisyFI|^2\right),
\end{equation}
which has also been employed in~\cite{xu_dynamic_2014, xu_regression_2015}.
All spectral coefficients of a segment~$\sFrameIdx$, i.e., $\sNoisyLogI$ for all frequency bins $\sFreqIdx$ for a given frame~$\sFrameIdx$, are stacked in a feature vector.

Given only the log-spectral coefficients, the \ac{DNN} needs to learn how to distinguish between speech and noise using the training data.
This is a potentially challenging task, as a large amount of different acoustic scenarios is required for training to match real conditions.
Hence, the approaches in~\cite{xu_dynamic_2014, kumar_speech_2016} sought to improve the robustness to unseen noise environments using non-\ac{ML} based noise \ac{PSD} estimators.
For this, the noisy log-spectral features given above have been extended by appending an estimate of the noise \ac{PSD}~\cite{xu_dynamic_2014, kumar_speech_2016}, which is also known as noise aware training~\cite{seltzer_investigation_2013}.
Similar to~\eqref{eq:LogPer}, the logarithmized estimate of the noise \ac{PSD} is given by
\begin{equation}
    \sVarNoiseFEstLog = \log\left(\sVarNoiseFEst\right).
\end{equation}
As a result, the feature vector for this set has twice the dimensionality as using only the log-spectral features.
In our experiments, the noise \ac{PSD} is estimated using the algorithm proposed in~\cite{gerkmann_noise_2011, gerkmann_unbiased_2012}, i.e., using the method described in Section~\ref{sec:NonMLAlgorithms:NoisePSD}.
For both features sets, a context of three past segments is added to this vector by appending the respective feature vectors to the end of the vector.
We do not add context from future segments to keep the algorithmic latency as low as for the non-\ac{ML} based enhancement scheme.

\subsection{\label{sec:DNN:Normalized}Proposed Normalized Features}

The main goal of the proposed normalized features is also to increase the robustness of \ac{DNN} based enhancement schemes to unseen noise conditions.
However, instead of appending the noise \acp{PSD} to the noisy input features, we incorporate the generalization of non-\ac{ML} based enhancement schemes by using the estimated noise \ac{PSD} as normalization term.
More specifically, we employ the logarithmized \emph{a priori} \ac{SNR}~$\sPriorSNRI^\text{(log)} = \log(\sPriorSNRI)$ and \emph{a posteriori} \ac{SNR}~$\sPostSNRI^\text{(log)} = \log(\sPostSNRI)$ as input features.
The \emph{a priori} \ac{SNR} and \emph{a posteriori} \ac{SNR} are defined as
\begin{align}
    \label{eq:PriorSNR}
    \sPriorSNRI &= \frac{\sVarSpeechF}{\sVarNoiseF}\\
    \label{eq:PosteriorSNR}
    \sPostSNRI &= \frac{|\sNoisyFI|^2}{\sVarNoiseF}.
\end{align}
The usage of the \emph{a priori} and \emph{a posteriori} \acp{SNR} is motivated by non-\ac{ML} based clean speech estimators, e.g.,~\cite{ephraim_speech_1984, ephraim_speech_1985, breithaupt_parameterized_2008, hendriks_dft-domain_2013}, where the quantities appear in the analytical solutions derived in a statistical framework.
The speech \ac{PSD}~$\sVarSpeechF$ and the noise \ac{PSD}~$\sVarNoiseF$ are estimated blindly from the noisy observation using the methods described in Section~\ref{sec:NonMLAlgorithms:NoisePSD}~\cite{gerkmann_noise_2011} and Section~\ref{sec:NonMLAlgorithms:SpeechPSD}~\cite{breithaupt_novel_2008}.
Both \acp{SNR} can be used separately or in combination by concatenating both in a single vector.
Note that the dimensionality of the features is the same as the noisy log-spectra if one of the \acp{SNR} is used as input.
In all considered cases, a temporal context of three previous segments is appended to the feature vectors.

In contrast to the non-normalized features in Section~\ref{sec:DNN:NonNormalized}, the \emph{a priori} and the \emph{a posteriori} \ac{SNR} exhibit the advantage that these features are scale-invariant.
As their value does not depend on the overall level, differently scaled training data results in identical normalized features as when the scaling is not varied.
To make the scale-invariance also available to a \ac{DNN} using non-normalized features, e.g., Section~\ref{sec:DNN:NonNormalized}, the training data has to reflect these gain variations.
This increases the variations in the training examples such that learning potentially becomes challenging.
Such gain variations do not increase the variability for the normalized features, which may allow to improve the enhancement.

\section{\label{sec:InstrumentalEvaluation}Instrumental Evaluation}

In this section, the algorithms described in Section~\ref{sec:NonMLAlgorithms} and Section~\ref{sec:DNN} are compared using instrumental measures.
Further, the \ac{OMLSA} proposed in~\cite{cohen_speech_2001, cohen_noise_2003} is used as a reference.
First, the audio material, the used parameters and instrumental measures are considered.
Afterwards, the results of the experimental analysis are presented which compares the properties of the normalized and the non-normalized input features in the \ac{DNN} framework.
Further, the computational complexity and the training convergence speed are considered.
In the last part of the instrumental evaluation, the performance of the evaluated algorithms is compared.

\subsection{\label{sec:Parameters}Audio Material, Parameters and Instrumental Measures}

For all algorithms, the \ac{STFT} uses 32~ms segments which overlap by 50~\%.
For the analysis step as well as the synthesis step a square-root Hann window is employed.
All signals have a sampling rate of 16~kHz.
For computing the features, the mirror spectrum is omitted such that the resulting dimension of the spectra is 257.
Correspondingly, the feature dimensionality of the noisy log-spectra~$\sNoisyLogI$, the \emph{a priori} \ac{SNR} $\sPriorSNRILog$ and the \emph{a posteriori} \ac{SNR} $\sPostSNRILog$ is $257 \times (3 + 1) = 1028$ including the context.
The dimensionality of the input features doubles to $2056$ for the combination of the \emph{a priori} \ac{SNR}~$\sPriorSNRI$ and the \emph{a posteriori}~\ac{SNR}, as well as, for the combination of the noisy log-spectra~$\sNoisyLogI$ and the logarithmized noise \ac{PSD}~$\sVarNoiseFEstLog$ as employed in~\cite{xu_dynamic_2014, kumar_speech_2016}.
The \ac{DNN}'s architecture comprises three hidden layers with \acp{ReLU}~\cite{nair_rectified_2010} as non-linearities and an output layer with sigmoidal activation functions.
The number of units in each hidden layers amounts to 1024 for both \ac{DNN} based approaches.
For the evaluations in this section, the minimum gain is set to $\sGainMin = -20~\text{dB}$ for all employed enhancement schemes.
For the non-\ac{ML} algorithms in Section~\ref{sec:NonMLAlgorithms} the parameters in the respective publication~\cite{breithaupt_novel_2008, gerkmann_noise_2011} are used.

The employed background noises are taken from a fixed pool of nine noise types.
It includes the babble noise and the factory~1 noise taken from the NOISEX-92 database~\cite{steeneken_description_1988}.
Further, a modulated version of NOISEX-92's pink and white noise are included as in~\cite{gerkmann_noise_2011}.
Additional noise types are taken from the freesound database \url{http://www.freesound.org}.
Among them are the sounds of an overpassing propeller plane (\url{https://freesound.org/s/115387/}), the interior of a passenger jet during flight (\url{https://freesound.org/s/188810/}), a vacuum cleaner (\url{https://freesound.org/s/67421/}) and a traffic noise (\url{https://freesound.org/s/75375/}).
Further, a two-talker babble noise is included which is generated using two read out stories taken from \url{https://www.vorleser.net}.
The two stories are read by a male and a female speaker, respectively, and are mixed at an \ac{SNR} of $0~\text{dB}$ after speech pauses have been removed.
The noise types are used to conduct cross-validation experiments where all noise types except one are included in the training set.
The training data of each cross-validation set are augmented by additionally including a highly non-stationary noise type which is generated from the noise snippets collected by~\cite{hu_corpus_nodate}.
The noise excerpts in this database are generally short and are, hence, concatenated multiple times in various orders to give a continuous noise signal.
Long noise excerpts are split into roughly 2~second long snippets during this generation.
This noise type is referred to as \ac{CSNE}.
The remaining unseen noise type is used for testing in the evaluations.

The speech material for training is taken from the TIMIT database~\cite{garofolo_timit_1993}.
For the training of the \ac{DNN} based enhancement schemes, a set of 1196~female and 1196~male sentences taken from the TIMIT training set is employed.
All sentences are embedded once in each noise type used for training at a random temporal position.
For the employed noise \ac{PSD} estimator, a two second initialization period is added at the beginning of each sentence  to avoid initialization artifacts during feature extraction.
This period is removed from the final features used for training.
However, a noise only period which amounts to $15~\%$ of the utterance length is included for each sentence in the training data.
To allow the \ac{DNN} to learn the effect of different \acp{SNR}, the sentences are embedded in the background noise at \acp{SNR} ranging from $-10~\text{dB}$ to $15~\text{dB}$.
The \ac{SNR} is randomly chosen for each sentence and also the scaling is randomly varied for each sentence by adjusting the peak level of the speech signal from $-26~\text{dB}$ and $-3~\text{dB}$.
These variations are included in the training data, to allow the \ac{DNN} based on the non-normalized features to learn a scale-independent function of the \ac{IRM}.

The parameters of the \ac{DNN} are adapted by minimizing the following optimization criterion
\begin{equation}
    \label{eq:CostFunction}
    \sCost = \sum_\sFrameIdx \sum_\sFreqIdx \left| \log\Big(\sIRMEst(\sFreqIdx, \sFrameIdx) + \sBias\Big) - \log\Big(\sIRM(\sFreqIdx, \sFrameIdx) + \sBias\Big) \right|^2.
\end{equation}
Here, the squared error of the logarithmized quantities is minimized which is motivated by the human loudness perception which approximately follows a logarithmic law.
Further, $\sBias$ is a bias term which is used to avoid that extremely low gains of the target \ac{IRM} are overly penalized by the cost function.
Here, $\sBias = 0.1$ is employed such that differences between the target \ac{IRM} and the \ac{DNN} output are treated as irrelevant if the target \ac{IRM} is below $-20~\text{dB}$.
The weights and biases of the layers are initialized using the Glorot method~\cite{glorot_understanding_2010}.
After the initialization, the weights are optimized using the AdaGrad approach~\cite{duchi_adaptive_2011} where the learning rate has been set to 0.005 while a batch size of 128~samples has been used.
The order of the training observations is randomized.
To avoid overfitting of the network, an early stopping scheme is employed where the training procedure is stopped if the error $J$ is not reduced by more than $1\,\%$ over 10~iterations on a validation set.
The validation set is constructed by randomly selecting 15~\% of the training set.

For testing, 128~sentences, 64~female and 64~male,  are taken from the TIMIT test set. 
Similar to the training, the clean speech sentences are embedded at random positions in the background.
All sentences are mixed at \acp{SNR} ranging from $-5~\text{dB}$ to $20~\text{dB}$ in $5~\text{dB}$ steps.
Furthermore, also here, an initialization period of two seconds is added to avoid initialization artifacts of the employed noise \ac{PSD} estimator~\cite{gerkmann_noise_2011}.
This period is omitted during the evaluation, i.e., the instrumental measures are only evaluated on the part that contains the embedded sentence.

For the comparison, \ac{PESQ}~\cite{noauthor_p.862:_2001} is used as an instrumental measure to evaluate the quality of the enhanced signals.
Generally, \ac{PESQ} improvement scores ($\Delta$\acs{PESQ}) are shown which are obtained by computing the difference between the \ac{PESQ} score of the enhanced and the noisy signal.
Further, \ac{STOI}~\cite{taal_algorithm_2011} is used to instrumentally predict the speech intelligibility of the enhanced signals.
In this evaluation, the \ac{STOI} scores are mapped to actual intelligibility scores, i.e., the percentage of words a human would correctly identify in a listening experiment.
As no mapping is available for the TIMIT database, the mapping function given for the IEEE sentences in~\cite{taal_algorithm_2011} is used.
Also here, improvements are computed ($\Delta$\acs{STOI}) which are obtained by subtracting the mapped speech intelligibility of the enhanced and the noisy signal.

\subsection{Analysis}

In this part, we give an analysis on the features proposed in Section~\ref{sec:DNN:NonNormalized} and Section~\ref{sec:DNN:Normalized}.
We demonstrate that the proposed normalized features in Section~\ref{sec:DNN:Normalized} are independent of scaling of the input signal.
Further, we show that the \ac{DNN} converges more quickly if the proposed features are employed.
In the last part of this section, the computational complexity of the various approaches is considered.

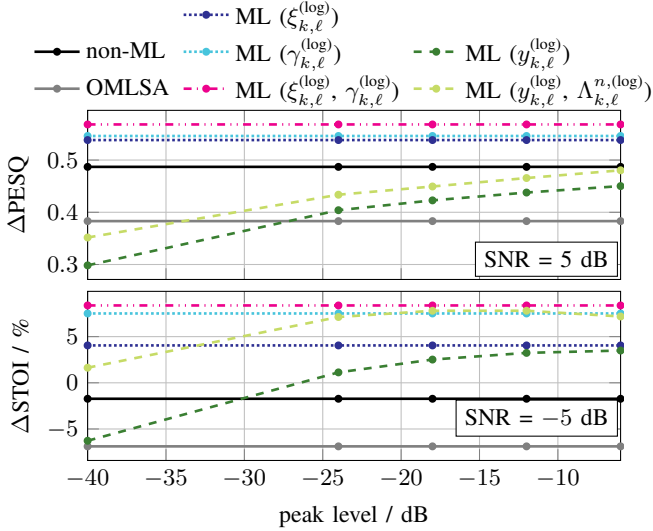
\begin{figure}[tb]
    \centering
    \hspace*{4ex}\begin{tikzpicture}
    \begin{customlegend}[
            legend entries={%
                { },
                \ac{ML} ($\sPriorSNRILog$),
                { },
                non-ML,
                \ac{ML} ($\sPostSNRILog$),
                \ac{ML} ($\sNoisyLogI$),
                OMLSA,
                {\ac{ML} ($\sPriorSNRILog$, $\sPostSNRILog$)},
                {\ac{ML} ($\sNoisyLogI$, $\sVarNoiseFLog$)},
            },
            legend columns={3},
            legend cell align={left},
            legend style={inner sep=1pt, draw=none, font=\small},
            /tikz/every even column/.append style={column sep=0.1cm},
        ]
        \addlegendimage{empty legend};
        \addlegendimage{log_aprior};
        \addlegendimage{empty legend};
        \addlegendimage{generic};
        \addlegendimage{log_apost};
        \addlegendimage{log_pery};
        \addlegendimage{omlsa};
        \addlegendimage{log_apost_aprior};
        \addlegendimage{log_pery_psdn};
    \end{customlegend}
\end{tikzpicture}\\
    \begin{tikzpicture}[]
    \begin{groupplot}[%
        scale only axis,
        group style={
            group size=1 by 2,
            x descriptions at=edge bottom,
            horizontal sep=4em,
            vertical sep=1ex,
        },
        width=0.8\linewidth,
        height=2.25cm,
        xlabel={peak level / dB},
        ylabel={},
        label style={font=\small},
        ticklabel style={font=\small},
        grid,
        xmax=-6,
        xmin=-40,
        ]

        \nextgroupplot[ylabel={$\Delta$\acs{PESQ}}]
        \edef\tmp{}
        \pgfplotsforeachungrouped \algo in {generic, omlsa, log_apost, log_aprior, log_apost_aprior, log_pery, log_pery_psdn} {%
            \eappto\tmp{%
                \noexpand\addplot[\algo] table[x=scaling, y=\algo] {Data/scaling_crossval8_avg_5_pesq_imp.tsv};
            }
        }

        \tmp
        \node[anchor=south east, draw, fill=white, font=\small] at($(rel axis cs:1,0) + (-1pt, 1pt)$) {\acs{SNR} = 5~dB};

        \edef\tmp{}

        \nextgroupplot[
            ylabel={$\Delta$\acs{STOI} / \%},
        ]
        \pgfplotsforeachungrouped \algo in {generic, omlsa, log_apost, log_aprior, log_apost_aprior, log_pery, log_pery_psdn} {%
            \eappto\tmp{%
                \noexpand\addplot[\algo] table[x=scaling, y=\algo] {Data/scaling_crossval8_avg_-5_stoi_map_imp.tsv};
            }
        }

        \tmp
        \node[anchor=south east, draw, fill=white, font=\small] at($(rel axis cs:1,0) + (-1pt, 9pt)$) {\acs{SNR} = $-5~\text{dB}$};
    \end{groupplot}
\end{tikzpicture}
    \caption{\label{fig:ScalingAvg}\acs{PESQ} and \acs{STOI} improvements for the considered enhancement algorithms in dependence of the peak level of the clean speech signal averaged over all noise types at an input \acs{SNR} of 0~dB.}
\end{figure}

To demonstrate the scale-invariance of the proposed features, the considered enhancement approaches are evaluated on speech material where the peak level of the speech utterances is varied systematically.
For this, we set the peak level of the speech utterances to $-6~\text{dB}$, $-12~\text{dB}$, $-18~\text{dB}$, $-24~\text{dB}$ and $-40~\text{dB}$.
The $-40~\text{dB}$ peak level has not been seen during training and can be considered an extreme case whereas the remaining levels are within the range of variations included in the training data.
For this evaluation, the \ac{SNR} of the input signals is fixed at $-5~\text{dB}$ for \ac{STOI} and $5~\text{dB}$ for \ac{PESQ}.
A lower \ac{SNR} is used for \ac{STOI} because the speech intelligibility reduces only considerably for \acp{SNR} lower than $0~\text{dB}$.
The results in terms of \ac{PESQ} and \ac{STOI} improvements are depicted in Fig.~\ref{fig:ScalingAvg}.
For this, the averages over all noise types excluding the \ac{CSNE}~\cite{hu_corpus_nodate} are computed.
The results show that the non-\ac{ML} based speech enhancement algorithms and the \ac{ML} based approaches based on the normalized features yield the same outcome independent of the scaling of the input signal.
Contrarily, the performance of the \ac{ML} based enhancement scheme using noisy log-spectra varies over the peak level of the input signal.
The same can be observed for the combination with the estimated noise \ac{PSD}.
This indicates that by using the normalized features, the \ac{ML} based algorithm does not depend on the overall level.
Contrarily, despite the efforts taken to increase the scale-independence during the training process, the non-normalized features result in scale-dependent results.

The convergence speed of the proposed features is measured using the number of epochs that have been required until the validation error converges.
Due to the cross-validation setup, nine models are trained for each feature type which allows to average the number of epochs required to train each model.
About 28 to 29 epochs are required on average if the non-normalized features are employed, whereas only 20 to 23 iterations are required for the normalized features.
This result provides evidence that the proposed normalized features simplify the training of the respective \acp{DNN}.

\begin{figure}[tb]
    \centering
    \begin{tikzpicture}[]
    \begin{axis}[%
        ybar,
        ymode=log,
        scale only axis,
        width=0.8\linewidth,
        height=2.25cm,
        xlabel style={yshift=2ex},
        ylabel={real-time factor},
        xtick={0,1,2,...,6},
        xticklabels={%
            \acs{OMLSA},
            non-\acs{ML},
            \acs{ML} ($\sNoisyLogI$),
            {\acs{ML} ($\sNoisyLogI, \sVarNoiseFLog$)},
            \acs{ML} ($\sPriorSNRILog$),
            \acs{ML} ($\sPostSNRILog$),
            {\acs{ML} ($\sPriorSNRILog, \sPostSNRILog$)},
        },
        xticklabel style={rotate=30, anchor=east, inner sep=0pt, yshift=-1.5ex, xshift=0.5ex, font=\footnotesize},
        ytick={1,2,4,8,16,32,64},
        yticklabels={$1$, $2$, $4$, $8$, $16$, $32$, $64$},
        ymin=1,
        ymax=40,
        ymajorgrids,
        point meta={exp(y)},
        nodes near coords={$\times\pgfmathprintnumber\pgfplotspointmeta$},
        nodes near coords style={Blue, font=\footnotesize, yshift=-0.5ex},
        ]

        \addplot[draw=Blue, thick, fill=RoyalBlue] table[x expr=\coordindex, y index=1, header=false] {Data/comp_complexity.tsv};
    \end{axis}
\end{tikzpicture}%
    \caption{\label{fig:CompComplexity}
        Computational complexity of the considered algorithms in terms of the real-time factor.
        This factor describes how many seconds of the audio material can be processed within a second in the real world.
        The number on top of the bar shows the actual value of the real-time factor.
    }
\end{figure}
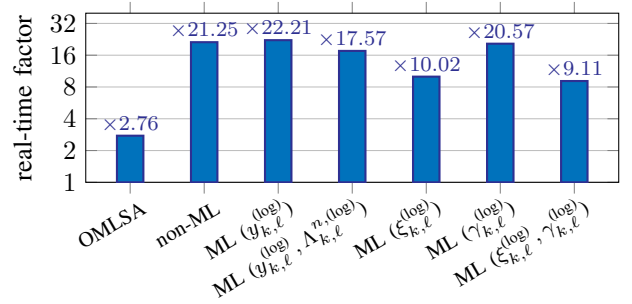

Last, the computational complexity of the considered algorithms is considered.
Fig.~\ref{fig:CompComplexity}, shows the processing speed of the various algorithms in terms of the real-time factor.
The factor describes how many seconds of the audio signal can be processed within a second in the real world.
Correspondingly, if the factor is larger than one, the algorithm processes the signals faster than real-time and if the factor is smaller than one, the processing is slower than real-time.
The algorithms have been evaluated on the CPU (Intel Core i7-5930K) of a desktop PC.
For this, their respective  Python or Matlab implementations have been used.
Fig.~\ref{fig:CompComplexity} shows that the \ac{OMLSA} runs slowest while the non-\ac{ML} described in Section~\ref{sec:NonMLAlgorithms} and the \ac{ML} based algorithms where the noisy log-spectra $\sNoisyLogI$ or the \emph{a posteriori} \ac{SNR} $\sPostSNRI$ are used as input feature run fastest.
On the used hardware, the quickest algorithms run roughly 20 times faster than real-time.
The \ac{OMLSA} is only about three times faster than real-time which may be explained by the fact that the Matlab implementation is run through Python which potentially introduces further processing overhead.
Using the \emph{a priori} \ac{SNR}~$\sPriorSNRILog$ instead of the \emph{a posteriori}~$\sPostSNRILog$ reduces the real-time factor to 10.
This is because, in comparison to $\sPostSNRILog$, the cepstral smoothing given in Section~\ref{sec:NonMLAlgorithms:SpeechPSD} needs to be additionally computed to obtain~$\sPriorSNRILog$.
Using both, the \emph{a priori} \ac{SNR}~$\sPriorSNRILog$ and the \emph{a posteriori} \ac{SNR}~$\sPostSNRILog$, as input features, the real-time factor further drops to 9.
Concatenating $\sNoisyLogI$ and $\sVarNoiseFLog$ is computationally more complex than using the $\sNoisyFI$ normalized by $\sVarNoiseF$, i.e., $\sPostSNRILog$.
For the concatenated features, the input dimensionality is twice as large as for the \emph{a posteriori} \ac{SNR} $\sPostSNRILog$ which results in the additional computational complexity.
From this, it is followed that the inclusion of the noise \ac{PSD} generally increases the computational complexity as expected.
Using $\sPostSNRILog$, the increase is only small whereas including the \emph{a priori} \ac{SNR} considerably increases the complexity.

\subsection{Comparisons}

\begin{figure*}[tb]
    \centering
    \begin{tikzpicture}
    \begin{customlegend}[
            legend entries={%
                OMLSA,
                non-ML,
                \ac{ML} ($\sNoisyLogI$),
                {\ac{ML} ($\sNoisyLogI$, $\sVarNoiseFLog$)},
                \ac{ML} ($\sPriorSNRILog$),
                \ac{ML} ($\sPostSNRILog$),
                {\ac{ML} ($\sPriorSNRILog$, $\sPostSNRILog$)},
            },
            legend columns={7},
            legend cell align={left},
            legend style={inner sep=1pt, draw=none, font=\small},
            /tikz/every even column/.append style={column sep=0.1cm},
        ]
        \addlegendimage{omlsa};
        \addlegendimage{generic};
        \addlegendimage{log_pery};
        \addlegendimage{log_pery_psdn};
        \addlegendimage{log_aprior};
        \addlegendimage{log_apost};
        \addlegendimage{log_apost_aprior};
    \end{customlegend}
\end{tikzpicture}\\
    \begin{tikzpicture}[
        axlabel/.style={fill=white, draw, inner sep=2pt},
    ]
    \begin{groupplot}[%
        scale only axis,
        group style={
            group size=5 by 2,
            x descriptions at=edge bottom,
            ylabels at=edge left,
            horizontal sep=5ex,
            vertical sep=1ex,
        },
        width=0.145\linewidth,
        height=2.25cm,
        ylabel={$\Delta$\acs{PESQ}},
        xmin=-5,
        xmax=20,
        xtick={-5,0,...,20},
        minor y tick num=1,
        grid,
        yminorgrids,
        ]
        \edef\tmp{}
        \pgfplotsforeachungrouped \noise/\noisehuman in {%
            mod_pink/mod. pink,
            mod_white/mod. white,
            overpassingplane/overpassing plane,
            aircraft-747-400-interior-midflight/aircraft interior,
            babble/babble,
            two_talker/two talker,
            factory1/factory,
            verynoisytrafficheavy/traffic,
            vacuum_cleaner/vacuum cleaner} {%
            \eappto\tmp{\noexpand\nextgroupplot}

            \eappto\tmp{\noexpand\node[font=\noexpand\footnotesize, anchor=south, draw, fill=white, inner sep=2pt] at($(rel axis cs:0.5,0) + (0pt,1pt)$) {\noisehuman};}

            \pgfplotsforeachungrouped \algo in {generic, omlsa, log_pery, log_pery_psdn, log_aprior, log_apost, log_apost_aprior} {%
                \eappto\tmp{%
                    \noexpand\addplot[\algo] table[x=snr, y=\algo] {Data/pesq_imp_crossval8_rnd_\noise.tsv};
                }
            }
        }

        \tmp

        \nextgroupplot[line width=1.25pt]

        \node[font=\footnotesize, anchor=south, draw, fill=white, inner sep=2pt] at($(rel axis cs:0.5,0) + (0pt,1pt)$) {\textbf{average}};

        \edef\tmp{}

        \pgfplotsforeachungrouped \algo in {generic, omlsa, log_pery, log_pery_psdn, log_aprior, log_apost, log_apost_aprior} {%
            \eappto\tmp{%
                \noexpand\addplot[\algo] table[x=snr, y=\algo] {Data/pesq_imp_crossval8_rnd_avg.tsv};
            }
        }

        \tmp
    \end{groupplot}
\end{tikzpicture}\\[-2ex]
    \begin{tikzpicture}[
        axlabel/.style={fill=white, draw, inner sep=2pt},
    ]
    \begin{groupplot}[%
        scale only axis,
        group style={
            group size=5 by 2,
            x descriptions at=edge bottom,
            ylabels at=edge left,
            horizontal sep=5ex,
            vertical sep=1ex,
        },
        width=0.145\linewidth,
        height=2.25cm,
        xlabel={SNR / dB},
        ylabel={$\Delta$\acs{STOI} / \%},
        xmin=-10,
        xmax=5,
        xtick={-10,-5,0,...,20},
        minor y tick num=1,
        grid,
        yminorgrids,
        ]
        \edef\tmp{}
        \pgfplotsforeachungrouped \noise/\noisehuman in {%
            mod_pink/mod. pink,
            mod_white/mod. white,
            overpassingplane/overpassing plane,
            aircraft-747-400-interior-midflight/aircraft interior,
            babble/babble,
            two_talker/two talker,
            factory1/factory,
            verynoisytrafficheavy/traffic} {%
            \eappto\tmp{\noexpand\nextgroupplot}

            \eappto\tmp{\noexpand\node[font=\noexpand\footnotesize, anchor=south east, draw, fill=white, inner sep=2pt] at($(rel axis cs:1,0) + (-1pt,1pt)$) {\noisehuman};}

            \pgfplotsforeachungrouped \algo in {generic, omlsa, log_pery, log_pery_psdn, log_aprior, log_apost, log_apost_aprior} {%
                \eappto\tmp{%
                    \noexpand\addplot[\algo] table[x=snr, y=\algo] {Data/stoi_map_imp_crossval8_rnd_\noise.tsv};
                }
            }
        }

        \tmp

        \nextgroupplot[ymin=-12, ymax=12.5]
        \edef\tmp{}

        \node[font=\footnotesize, anchor=south east, draw, fill=white, inner sep=2pt] at($(rel axis cs:1,0) + (-1pt,1pt)$) {vacuum cleaner};

        \pgfplotsforeachungrouped \algo in {generic, omlsa, log_pery, log_pery_psdn, log_aprior, log_apost, log_apost_aprior} {%
            \eappto\tmp{%
                \noexpand\addplot[\algo] table[x=snr, y=\algo] {Data/stoi_map_imp_crossval8_rnd_vacuum_cleaner.tsv};
            }
        }
        \tmp

        \nextgroupplot[line width=1.25pt]
        \edef\tmp{}

        \node[font=\footnotesize, anchor=south east, draw, fill=white, inner sep=2pt] at($(rel axis cs:1,0) + (-1pt,1pt)$) {\textbf{average}};

        \pgfplotsforeachungrouped \algo in {generic, omlsa, log_pery, log_pery_psdn, log_aprior, log_apost, log_apost_aprior} {%
            \eappto\tmp{%
                \noexpand\addplot[\algo] table[x=snr, y=\algo] {Data/stoi_map_imp_crossval8_rnd_avg.tsv};
            }
        }
        \tmp

    \end{groupplot}
\end{tikzpicture}
    \caption{
        \label{fig:CrossVal8PESQSTOI}\acs{PESQ} and \acs{STOI} improvements that results for the considered enhancement algorithms in dependence of the background noise type and the \acs{SNR}.
        The \acs{ML} based algorithms are always trained on \acs{CSNE}~\cite{hu_corpus_nodate} and the noise types not given in the respective plot, i.e., the background noise is unseen in all cases.
    }
\end{figure*}
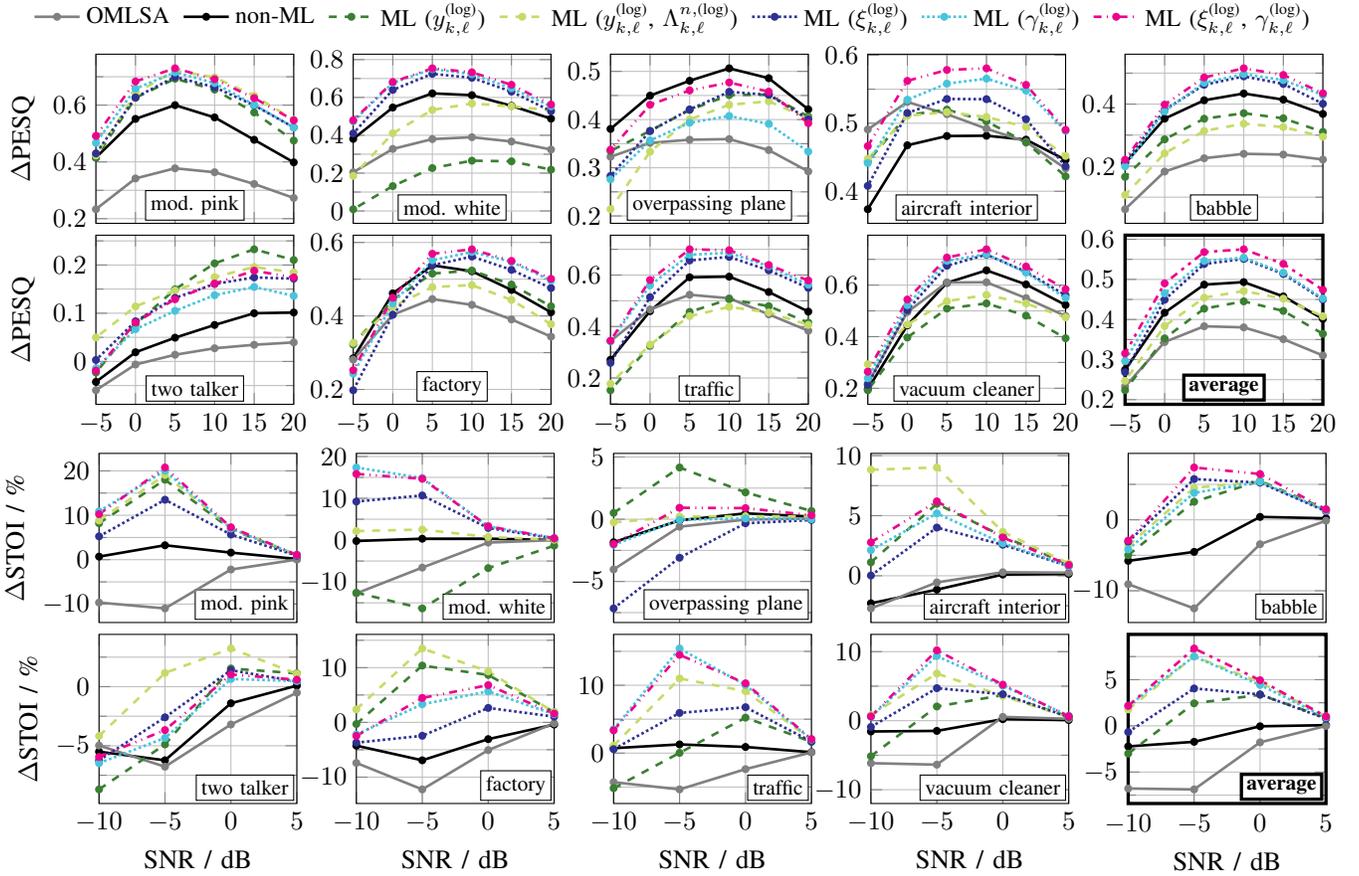

The following results show the outcome of the cross-validation procedure and are used to compare the enhancement algorithms used in this paper.
For these experiments, the peak level of the 128~TIMIT sentences used for testing is randomly varied between $-6~\text{dB}$ and $-26~\text{dB}$ which is similar to the range used for training.
Fig.~\ref{fig:CrossVal8PESQSTOI} depicts the results.

For both instrumental measures, first the performance of the non-\ac{ML} based approach is considered.
For the aircraft interior noise, the \ac{OMLSA} achieves higher \ac{PESQ} scores in low \acp{SNR}.
This is, however, an exception as for the remaining noise types, especially the nonstationary ones such as babble noise or the amplitude modulated versions of the pink and white noise, the performance of the employed non-\ac{ML} enhancement approach is higher than for the \ac{OMLSA}.
In terms of the speech intelligibility predicted by \ac{STOI}, both non-\ac{ML} approaches have either little effect or reduce the intelligibility.
In all cases, however, $\Delta$\ac{STOI} is higher for the approach described in Section~\ref{sec:NonMLAlgorithms} than for the \ac{OMLSA}.
Consequently, the algorithm described in Section~\ref{sec:NonMLAlgorithms} generally outperforms the \ac{OMLSA}~\cite{cohen_speech_2001, cohen_noise_2003}.

The speech intelligibility predicted by \ac{STOI} is generally higher for the \ac{ML} based algorithms than for the non-\ac{ML} approaches.
In factory noise and the aircraft interior noise, \ac{STOI} predicts a higher speech intelligibility for the non-normalized features.
The same is true for the overpassing plane and the two-talker noise.
Here, however, $\Delta$\ac{STOI} is generally smaller compared to the other noise types.
Among the non-normalized features, $\Delta$\ac{STOI} is generally higher for the combination of the noisy log-spectra~$\sNoisyLogI$ and the estimated noise \ac{PSD}~$\sVarNoiseFLog$.
For the remaining noise types, however, the proposed normalized features yield similar or higher \ac{STOI} improvements than the non-normalized features.
Comparing the normalized feature sets amongst each other shows that using only the \emph{a priori} \ac{SNR}~$\sPriorSNRILog$ often results in the lowest scores.
Contrarily, the combination of the \emph{a priori} \ac{SNR}~$\sPriorSNRILog$ and the \emph{a posteriori} \ac{SNR}~$\sPostSNRILog$ generally yields the highest scores.
In many cases, using only the \emph{a posteriori} \ac{SNR}~$\sPostSNRILog$ yields scores similar to the combination.
For cases, where the computational complexity plays an important role this feature type is thus a considerable alternative.

The \ac{PESQ} improvements for the \ac{ML} based algorithms indicate a clear preference for the proposed normalized features.
Only for the two talker noise, the \ac{PESQ} improvements obtained for the non-normalized features are higher than for the normalized features.
However, as basically all the considered enhancement algorithms struggle in this noise type, the gains of 0.05~points are rather small and therefore negligible.
For most of the remaining noise types, the performance of the non-normalized features predicted by \ac{PESQ} is between the \ac{OMLSA} and the non-\ac{ML} approach described in Section~\ref{sec:NonMLAlgorithms}.
Except for the modulated white noise, \ac{PESQ} does not indicate considerable advantages if an estimate of the noise \ac{PSD}~$\sVarNoiseFLog$ is appended to the noisy log-spectra~$\sNoisyLogI$.
This changes if the normalized features are used.
Using these features, the performance of the \ac{ML} approach is more robust and, often, both non-\ac{ML} approaches are outperformed.
Again, the combination of the \emph{a priori} \ac{SNR}~$\sPriorSNRILog$ and \emph{a posteriori} \ac{SNR}~$\sPostSNRILog$ yields the highest scores in most noise types.
Also here, using the \emph{a posteriori} \ac{SNR}~$\sPostSNRILog$ without the \emph{a priori} \ac{SNR}~$\sPriorSNRILog$ yields similar results as the combination of both.
Consequently, it is possible to benefit from the advantages of the normalized features without severely increasing the computational complexity.
Further, as this feature type has the same dimensionality as the noisy log-spectra, this demonstrates the importance of the normalized features on the generalization of \ac{ML} based enhancement schemes.

\section{Subjective Evaluation}
\label{sec:SubjectiveEvaluation}

Instrumental measures such as \ac{PESQ} give an indication on how the quality of the processed signals would be judged by humans.
Still, as such measures cannot perfectly model human perception, we verify the instrumental results in Section~\ref{sec:InstrumentalEvaluation} using subjective evaluation tests.
Here, a \ac{MUSHRA}~\cite{noauthor_bs.1534-3:_2015} is employed to compare the algorithms described in Section~\ref{sec:NonMLAlgorithms} and Section~\ref{sec:DNN}.
First, the audio material, parameters and evaluation are explained and, after that, the results are discussed.

\subsection{Audio Material, Parameters and Setup}

For this experiment, a sentence of a male and a female speaker is embedded in factory~1 noise and traffic noise at an \ac{SNR} of 5~dB.
The noisy signals are processed by the speech enhancement schemes described in Section~\ref{sec:NonMLAlgorithms} and Section~\ref{sec:DNN}.
The \ac{ML} based algorithm is included once using the noisy log-spectra as features $\sNoisyLogI$ and once using the combination of \emph{a priori} \ac{SNR}~$\sPriorSNRILog$ and \emph{a posteriori} \ac{SNR}~$\sPostSNRILog$.
For this experiment, the \ac{CSNE}~\cite{hu_corpus_nodate} and the two talker noise are excluded from the noise type pool such that eight noise types remain.
We train the \ac{DNN} once on a set which includes mod.\ pink noise, mod.\ white noise, factory 1 noise and traffic noise.
Note that this includes the traffic and factory noise which is also used for testing, i.e., this corresponds to a seen condition.
For this condition, it is ensured that the noise realizations used for the training are not reused for testing.
Therefore, only the first $120~\text{s}$ of the noise types are used while the last $120~\text{s}$ are used to embed the sentences for the listening experiment.
The algorithms have also been evaluated in an unseen condition where all noise types are included in the training set except the one used for evaluation.
Here, the full length of the training noise is utilized.
For each sentence embedded in the training noise type, the peak level is varied between $-26~\text{dB}$ and $-6~\text{dB}$, while the \ac{SNR} is chosen between $-5~\text{dB}$ and $15~\text{dB}$.
The minimum gain is set to $\sGainMin = -15~\text{dB}$ in this experiment.

In each trial of the experiment, the participants compared six stimuli.
In addition to the processed signals, the noisy signal is included and a reference signal is presented where the speech signal and the background noise are mixed at an \ac{SNR} of $20~\text{dB}$.
Lastly, a low quality anchor is added where the speech signal is low pass filtered at $2~\text{kHz}$ and mixed at an \ac{SNR} of $-5~\text{dB}$.
This signal is enhanced using a non-\ac{ML} based enhancement algorithm where the noise \ac{PSD} is estimated using~\cite{gerkmann_noise_2011} while the speech \ac{PSD} is obtained using the decision-directed approach~\cite{ephraim_speech_1984}.
The smoothing constant is set to $0.9$ and the signal is enhanced using the Wiener filter where a more aggressive lower limit of $-20~\text{dB}$ is employed.
This results in an anchor signal with very poor quality due to many musical tone artifacts and strong speech distortions.
The audio examples used for the listening experiment are available under \url{https://www.inf.uni-hamburg.de/en/inst/ab/sp/publications/tasl2017-dnn-rr}.

A total of 11~subjects with age in the range of 24 to 38 years who are not familiar with single-channel signal processing have participated in the \ac{MUSHRA}.
The experiment took place in a quiet office.
The diotic signals were presented via Beyerdynamic DT-770 Pro 250 Ohm headphones attached to an RME Fireface UFX+ sound card.
All signals were normalized in amplitude.
The test consisted of two phases.
First, the participants were asked to complete a training phase to familiarize with the presented sounds and to adjust the volume to a comfortable level.
For this, a subset of the processed signals was presented.
In the second part of the experiment, the participants were asked to rate the signals according to their overall preference on a scale from 0 to 100, where 0 was labeled with “bad” and 100 with “excellent”.
The order of the presentation of algorithms and conditions were randomized between all subjects.

\begin{figure}[tb]
    \centering
    \begin{tikzpicture}[
        skip loop/.style={to path={-- ++(0,#1) -| (\tikztotarget)}},
        nonsignificant/.style={thick, RoyalBlue},
        evaltype/.style={fill=white,
                         draw,
                         anchor=north west,
                         inner sep=2pt},
        faketitle/.style={font=\small, anchor=south, yshift=-0.3ex, anchor=south},
        noisetitle/.style={draw, fill=white, inner sep=2pt, anchor=south east},
    ]
    \begin{groupplot}[%
        scale only axis,
        group style={
            group size=2 by 2,
            x descriptions at=edge bottom,
            y descriptions at=edge left,
            horizontal sep=1ex,
            vertical sep=1ex,
        },
        width=0.39\linewidth,
        height=2.25cm,
        ylabel={Rating},
        boxplot/draw direction=y,
        boxplot/every median/.style={very thick},
        boxplot/every box/.append style={fill=white},
        boxplot/box extend=0.5,
        xtick={1,2,...,6},
        xticklabels={%
            anchor,
            noisy,
            non-ML,
            {\acs{ML} non-norm.},
            {\acs{ML} norm.},
        },
        xticklabel style={inner sep=1pt, rotate=35, anchor=north east, font=\small},
        ytick={0,25,50,75,100},
        ymin=-10,
        ymax=115,
        grid,
        ]
        \edef\tmp{}
        \pgfplotsforeachungrouped \eval in {seen, crossval7} {%
            \pgfplotsforeachungrouped \noise in {factory1, traffic} {%
                \pgfmathsetmacro{\i}{0}

                \eappto\tmp{\noexpand\nextgroupplot}

                \pgfplotsforeachungrouped \stimulus in {anchor, noisy, generic, log_noisy, log_spp_tcs_prio} {%
                    \pgfmathtruncatemacro{\i}{\i + 1}
                    \eappto\tmp{\noexpand\addplot[boxplot, boxplot/every box/.style={fill=RoyalBlue!25}, boxplot/draw position=\i, mark=x] table[y=\stimulus] {Data/subjective_\eval_\noise.tsv}
                        coordinate (upper_\stimulus_\eval_\noise) at (boxplot box cs: \noexpand\boxplotvalue{upper whisker}, 0.5)
                        coordinate (lower_\stimulus_\eval_\noise) at (boxplot box cs: \noexpand\boxplotvalue{lower whisker}, 0.5);
                    }

                }
            }
        }

        \tmp
    \end{groupplot}

    \draw[nonsignificant] (upper_anchor_seen_factory1) edge[latex-latex, skip loop=0.4] (upper_noisy_seen_factory1);
    \draw[nonsignificant] (upper_log_spp_tcs_prio_seen_factory1) edge[latex-latex, skip loop=0.4] (upper_log_noisy_seen_factory1);

    \draw[nonsignificant] (lower_anchor_seen_traffic) edge[latex-latex, skip loop=-0.3] (lower_noisy_seen_traffic);
    \draw[nonsignificant] (upper_log_spp_tcs_prio_seen_traffic) edge[latex-latex, skip loop=0.3] (upper_generic_seen_traffic);
    \draw[nonsignificant] (lower_log_spp_tcs_prio_seen_traffic) edge[latex-latex, skip loop=-0.4] ($(lower_log_noisy_seen_traffic) + (2pt, 0)$);
    \draw[nonsignificant] (lower_generic_seen_traffic) edge[latex-latex, skip loop=-0.3] ($(lower_log_noisy_seen_traffic) + (-2pt, 0)$);

    \draw[nonsignificant] (upper_log_noisy_crossval7_factory1) edge[latex-latex, skip loop=0.4] (upper_generic_crossval7_factory1);

    \draw[nonsignificant] (upper_log_spp_tcs_prio_crossval7_traffic) edge[latex-latex, skip loop=0.4] (upper_generic_crossval7_traffic);

    \node[evaltype] at($(group c1r1.north west) + (1pt, -1pt)$) {seen};
    \node[noisetitle] at($(group c1r1.south east) + (-1pt, 1pt)$) {factory 1};
    \node[evaltype] at($(group c2r1.north west) + (1pt, -1pt)$) {seen};
    \node[noisetitle] at($(group c2r1.south east) + (-1pt, 1pt)$) {traffic};

    \node[evaltype] at($(group c1r2.north west) + (1pt, -1pt)$) {unseen};
    \node[noisetitle] at($(group c1r2.south east) + (-1pt, 1pt)$) {factory 1};
    \node[evaltype] at($(group c2r2.north west) + (1pt, -1pt)$) {unseen};
    \node[noisetitle] at($(group c2r2.south east) + (-1pt, 1pt)$) {traffic};

\end{tikzpicture}
    \vspace*{-3ex}
    \caption{\label{fig:SubjectiveEvaluation}Box plots for the subjective rating of different enhancement schemes.
        The left column shows the results for factory~1 noise and the right column for traffic noise as test signals.
        The rows show different training strategies.
        The linking lines show pairings that are \emph{not} identified as statistically significant by the post-hoc tests.
    }
\end{figure}
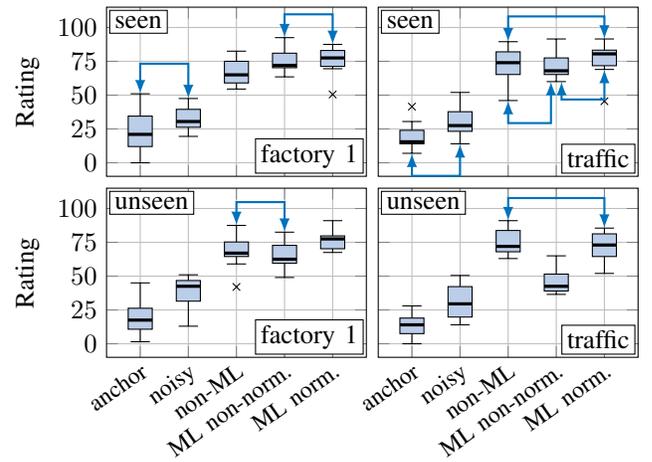

\subsection{Results}

For the evaluation, we average the ratings over the two speakers for each tested scenario.
Further, the results are validated using a statistical analysis.
For each acoustic scenario, a repeated measures \ac{ANOVA}~\cite{field_disocvering_2009} is performed to test if the factor “enhancement algorithm” has a significant effect on the participants' rating.
For this, we employ a significance level of 5~\% for all statistical tests.
For each acoustic scenario, we validated that the residuals of the general linear model fitted during the process of the repeated measures \ac{ANOVA} are normally distributed using the Shapiro-Wilk test~\cite{shapiro_analysis_1965}.
The sphericity assumption has been validated using Mauchly's test~\cite{mauchly_significance_1940} and a Greenhouse-Geisser correction~\cite{greenhouse_methods_1959} is employed in cases where it has been violated.
In all acoustic scenarios, the enhancement algorithms have a statistically significant effect on ratings.
Hence, post-hoc tests are used to identify the sources of significance.
For this, matched pair $t$-tests with a Bonferroni-Holm~\cite{holm_simple_1979} correction are employed to account for the error inflation.
The results are shown in Fig.~\ref{fig:SubjectiveEvaluation} where the ratings that are statistically \emph{not} significantly different are indicated by linking lines.

All listeners were able to correctly identify the hidden reference and assigned the highest score to it.
The anchor signal and the noisy signal were assigned the lowest scores in most of the cases.
For the seen conditions, all enhancement schemes have been rated similar in traffic noise, while in factory noise, both \ac{ML} based speech enhancement schemes yield slightly better results than the non-\ac{ML} based algorithm.
For the unseen conditions, the ratings for the \ac{ML} based approach only using the non-normalized noisy log-spectra as features drop while the ratings for the proposed normalized features remain high.
Additionally, the proposed features show slightly higher ratings in comparison to the non-\ac{ML} enhancement scheme in factory noise.
The statistical evaluation confirms that the highlighted differences are statistically significant.

\section{Conclusions}

In this paper, we propose features for \ac{ML} based speech enhancement which incorporate non-\ac{ML} based estimates of the speech and noise \ac{PSD}.
The goal is to improve the robustness of \ac{ML} based enhancement scheme towards unseen noise conditions.
In contrast to the already existing noise aware training~\cite{seltzer_investigation_2013, xu_dynamic_2014, xu_regression_2015, kumar_speech_2016}, the noise \ac{PSD} is not appended but used as a normalizing term.
This results in the \emph{a priori} \ac{SNR} and the \emph{a posteriori} \ac{SNR} which exhibit the advantageous property of being scale-invariant.
For the noisy log-spectra, the performance of the \ac{ML} based enhancement scheme in terms of \ac{PESQ} is low in unseen noise conditions.
Appending an estimate of the noise \ac{PSD} has only a little impact on the performance in \ac{PESQ} while the intelligibility predicted by \ac{STOI} increases.
Using the proposed normalized features, however, the performance of the \ac{ML} based enhancement scheme is generally higher as for the compared algorithms in both instrumental measures.
This is supported by the \ac{MUSHRA} based listening experiments where, in unseen noise conditions, the proposed combination was significantly preferred over the \ac{ML} based enhancement scheme using only the log-spectra of the noisy observations.
Audio examples are available under \url{https://www.inf.uni-hamburg.de/en/inst/ab/sp/publications/tasl2017-dnn-rr}.
Feed-forward networks clearly benefit from the proposed normalized features, but their effect on other architectures such as recurrent neural networks or convolutional networks remains a question for future research.

\ifCLASSOPTIONcaptionsoff
  \newpage
\fi

\bibliographystyle{IEEEtran}

\bibliography{Bibliography/Journal2017.bib,Bibliography/Control.bib}

\end{document}